\documentclass[conference]{IEEEtran}
\IEEEoverridecommandlockouts
\usepackage{cite}
\usepackage{amsmath,amssymb,amsfonts}
\usepackage{amssymb}
\usepackage{algorithmic}
\usepackage{graphicx}
\usepackage{textcomp}
\usepackage{xcolor}
\usepackage{cite}
\usepackage{amsmath,amssymb,amsfonts}
\usepackage{algorithmic}
\usepackage{graphicx}
\usepackage{textcomp}
\usepackage{times}
\usepackage{xcolor}
\usepackage{float}
\usepackage{algorithm}
\usepackage{multirow}

\usepackage{mathtools}
\usepackage{mathtools, nccmath}
\usepackage[font=small,skip=0pt]{caption}
\usepackage{fancyhdr}
\usepackage{hyperref}
\usepackage{makecell}
\usepackage{tikz}
\def\checkmark{\tikz\fill[scale=0.4](0,.35) -- (.25,0) -- (1,.7) -- (.25,.15) -- cycle;}
\usepackage{pifont}
\newcommand{\xmark}{\text{\ding{55}}}

\def\BibTeX{{\rm B\kern-.05em{\sc i\kern-.025em b}\kern-.08em
    T\kern-.1667em\lower.7ex\hbox{E}\kern-.125emX}}

\begin{document}

\title{Generalised Mathematical Formulations for Non-Linear Optimized Scheduling \\
\thanks{This research work has been done in the field of 5G Advanced and Beyond Scheduling for network slicing by Sharvari Ravindran, Saptarshi Chaudhuri, Jyotsna Bapat, and Debabrata Das, IIIT Bangalore}
}

\author{\IEEEauthorblockN{Sharvari Ravindran, Saptarshi Chaudhuri, Jyotsna Bapat, and Debabrata Das}
\IEEEauthorblockA{\textit{Networking and Communication Research Lab} \\
\textit{International Institute of Information Technology, Bangalore, India}\\
Sharvari.R@iiitb.ac.in, saptarshi.chauduri@iiitb.org, jbapat@iiitb.ac.in, ddas@iiitb.ac.in
}
}


\maketitle

\begin{abstract}
In practice, most of the optimization problems are non-linear requiring certain interactive solutions and approaches to model. In 5G Advanced and Beyond network slicing, mathematically modeling the users, type of service distributions and it’s adaptive SLAs are complex due to several dependencies. To facilitate the above, in this paper, we present novel Non-linear mathematical formulations and results that will form the base to achieve Optimized Scheduling.
\newline
\end{abstract}

\section{Result 1}

\textbf{Result 1}: \textit{Estimation with graphical interpretations of the Karush Kuhn Tucker (KKT) Lagrangian multipliers} $\mu_{i} \geq 0 \; \forall \; \exists \; i$.

For a generalized non-linear optimization problem, Karush Kuhn Tucker (KKT) [1][2] conditions are necessary criterion conditioned on certain regularity estimates. It starts off with formulating a Lagrangian as a function of the objectives and constraints brought together through multipliers. The objective is a quick formulation to analyze the KKT Lagrangian multipliers and its value. To estimate this, three cases are analyzed,
\begin{enumerate}
\item $\mathbf{Case \; 1}$: $\mathbf{O}$ is a maximization problem with $\mathbf{C} \geq 0$.
\item $\mathbf{Case \; 2}$: $\mathbf{O}$ is a minimization problem with $\mathbf{C} \leq 0$.
\item $\mathbf{Case \; 3}$: $\mathbf{O}$ is a max/min problem with $\mathbf{C} \lesseqgtr 0$.
\end{enumerate}
$\mathbf{Case \; 1}$: 
Consider the following Optimization problem (OP),
\begin{equation}
\mathcal{P} \Rightarrow
\begin{cases}
\mathbf{max} \; \mathbf{O}x(z) \Rightarrow \mathbf{O}x(z) > 0 \; \forall \; z, x=1,...,m \\
\mathbf{C}y(z) \geq 0  \Rightarrow \mathbf{C}y(z) > 0 \; \forall \; z, y=1,...,n \\
\end{cases}
\end{equation}

where $x, y$ denotes the index of the objective function and constraints and $z$ is the optimization variable whose solution is to be learnt. Since $(1)$ is a maximization problem, formulating the KKT Lagrangian with respect to $\mathbf{O}$ and $\mathbf{C}$ independently. Re-expressing $\mathbf{C}y(z)$ as a typical constraint formulation, i.e., $- \mathbf{C}y(z) \leq 0$,
\begin{equation}
\mathcal{L}(z, \mu_{y}) = \mathbf{O}x(z) - \mu_{y} (-\mathbf{C}y(z)) \; \forall \; x, y
\end{equation} 

where $\mu_{y} \geq 0$ is the multiplier that satisfies the following regularity condition of $\mathcal{L}(z, \mu_{y})  = 0$.
\begin{equation}
\underbrace{\triangledown_{z} \mathbf{O}x(z)}_\text{$> 0$} - \mu_{y} \triangledown_{z} (-\underbrace{\mathbf{C}y(z)}_\text{$> 0$}) = 0
\end{equation}

\begin{figure}[h!]
\begin{center}
\includegraphics[scale = 0.24]{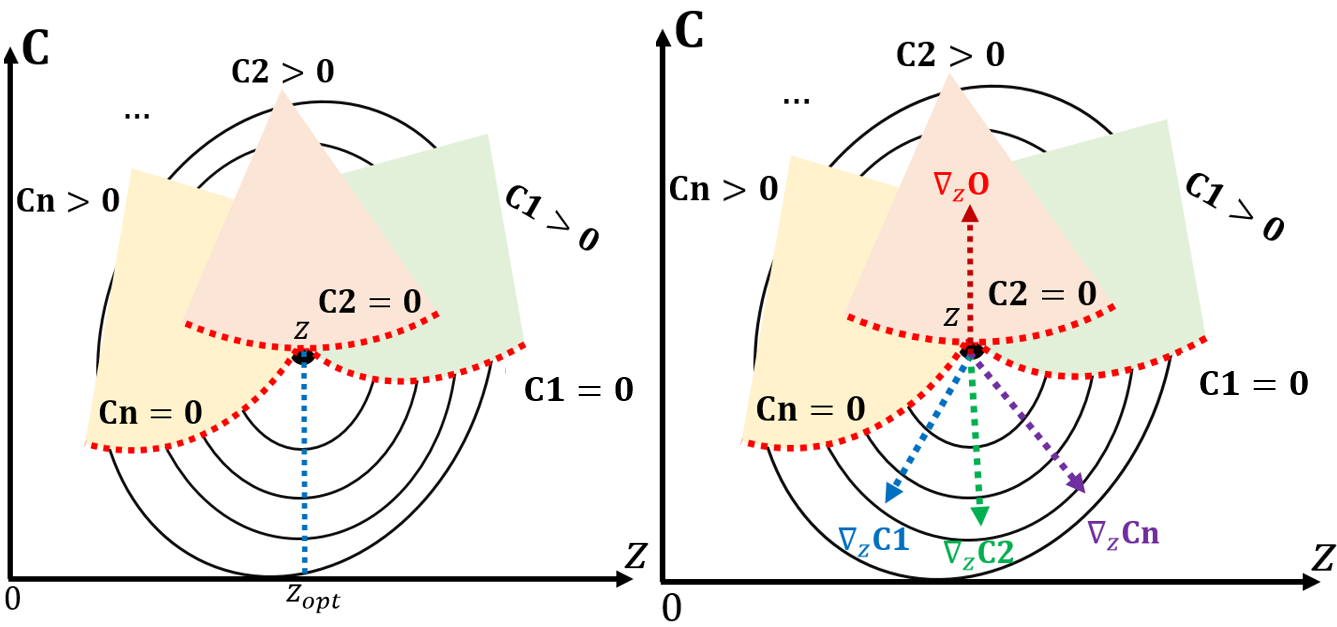}
\end{center}
\caption{$\mathbf{Case \; 1}$: Graphical interpretation of KKT Lagrangian multiplier estimation for maximization OPs}
\vspace{-5mm}
\end{figure}

\begin{figure}[h!]
\begin{center}
\includegraphics[scale = 0.25]{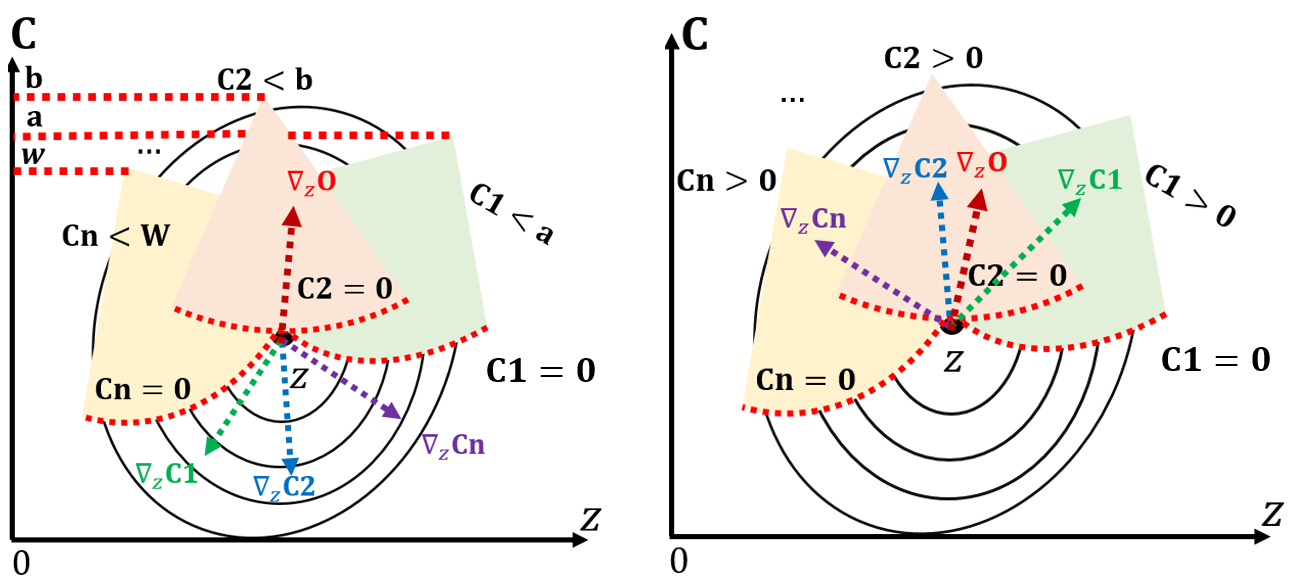}
\end{center}
\caption{Graphical interpretation of KKT Lagrangian multiplier estimation for minimization OP (objective) (a) $\mathbf{Case \; 2}$: $\mathbf{C}y < 0$, (b) $\mathbf{Case \; 3}$: $\mathbf{C}y > 0 \; \forall \; y$}
\vspace{-5mm}
\end{figure}

\begin{figure}[h!]
\begin{center}
\includegraphics[scale = 0.24]{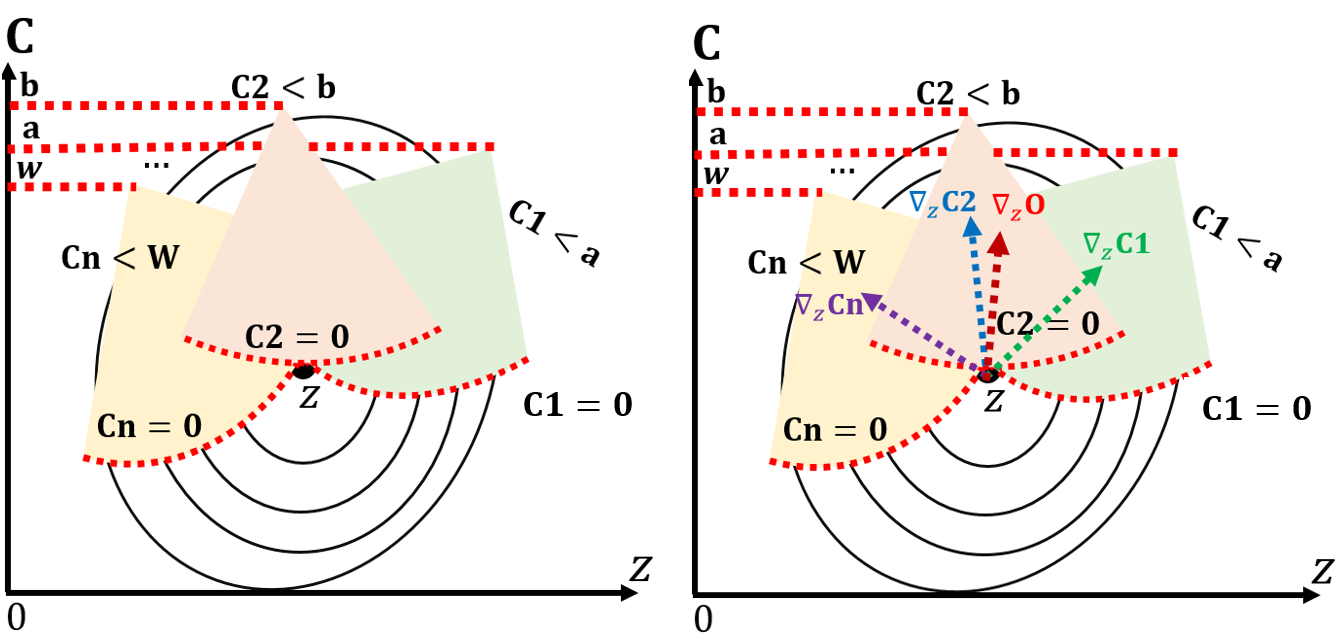}
\end{center}
\caption{$\mathbf{Case \; 3}$: Graphical interpretation of KKT Lagrangian multiplier estimation for maximization OP (objective)}
\vspace{-5mm}
\end{figure}

\begin{table*}[h!]
\caption{KKT Lagrangian multipliers for special cases}
\begin{center} 
\begin{tabular}{|p{4.5cm}||p{1cm}|p{1cm}|p{1cm}||p{1cm}|p{1cm}|p{1cm}|}
 \hline
$\mathbf{Criteria}$ & \multicolumn{3}{|c|}{$\mu_{y} > 0$} &  \multicolumn{3}{|c|}{$\mu_{y} = 0$}  \\
 \hline
 $\mathbf{Objective/Constraints}$ & $\mathbf{Case \; 1}$ &  $\mathbf{Case \; 2}$ &  $\mathbf{Case \; 3}$ & $\mathbf{Case \; 1}$ &  $\mathbf{Case \; 2}$ &  $\mathbf{Case \; 3}$ \\
 \hline
 $\triangledown_{z} \mathbf{O}x(z) > 0, \triangledown_{z} \mathbf{C}y(z) < 0 \; \forall \; z$ & $\checkmark$ & \checkmark & $\xmark$ & $\xmark$ & $\xmark$ & $\checkmark$ \\
 \hline
 $\triangledown_{z} \mathbf{O}x(z) < 0, \triangledown_{z} \mathbf{C}y(z) < 0 \; \forall \; z$ & $\xmark$ & $\xmark$ & $\checkmark$ & $\checkmark$ & $\checkmark$ & $\xmark$ \\
 \hline
 $\triangledown_{z} \mathbf{O}x(z) < 0, \triangledown_{z} \mathbf{C}y(z) > 0 \; \forall \; z$ & $\checkmark$ & \checkmark & $\xmark$ & $\xmark$ & $\xmark$ & $\checkmark$ \\
\hline
 \end{tabular}
 \end{center}
\end{table*}

\begin{figure}[h!]
\begin{center}
\includegraphics[scale = 0.25]{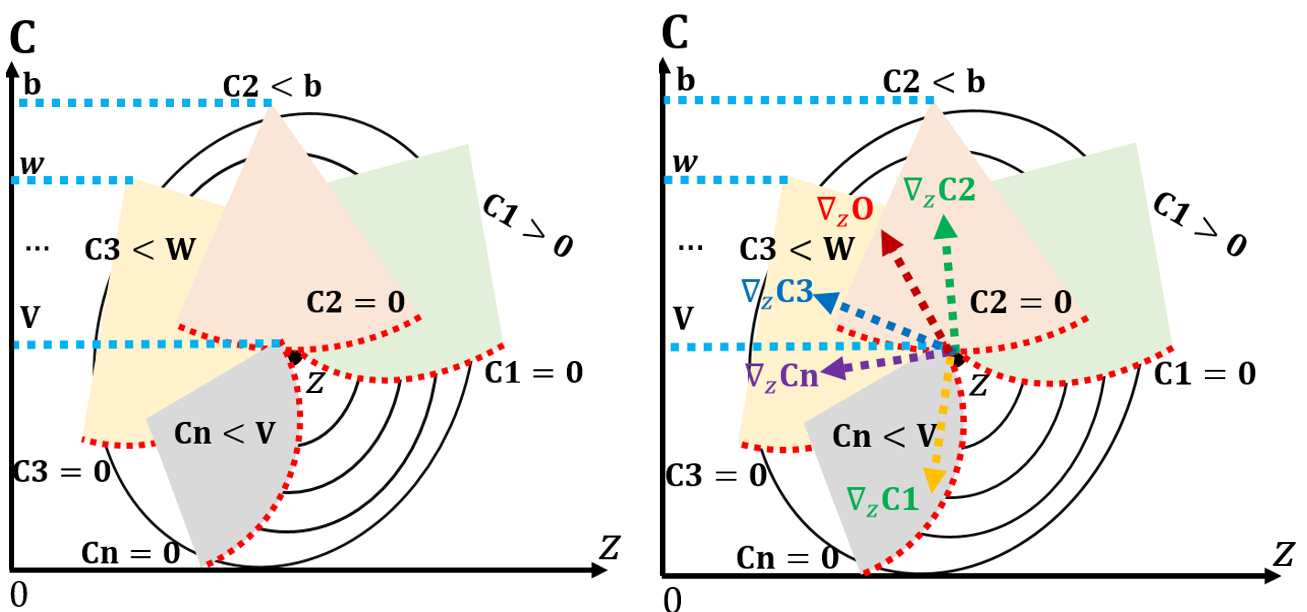}
\end{center}
\caption{$\mathbf{Extended \; result \; 1}$: Graphical interpretation of KKT Lagrangian multiplier estimation for maximization OP (objective)}
\vspace{-5mm}
\end{figure}
\begin{figure}[h!]
\begin{center}
\includegraphics[scale = 0.26]{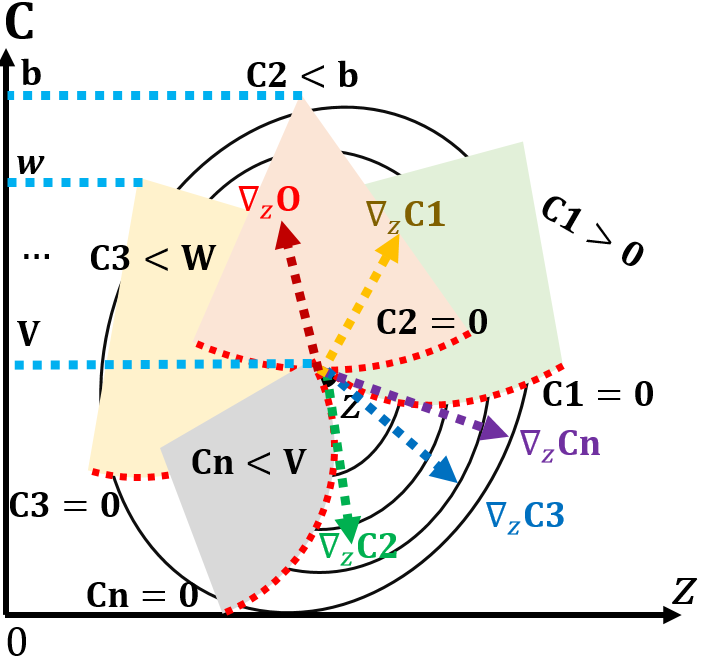}
\end{center}
\caption{$\mathbf{Extended \; result \; 2}$: Graphical interpretation of KKT Lagrangian multiplier estimation for minimization OP (objective)}
\vspace{-5mm}
\end{figure}

\begin{equation}
\mu_{y} = \frac{\triangledown_{z} \mathbf{O}x(z)}{\underbrace{- \triangledown_{z} \mathbf{C}y(z)}_\text{$\triangledown_{z} \mathbf{C}y(z)$}}
\end{equation} 

Eqn. $(4)$, i.e., $\mu_{y} < 0$ contradicts the definition and assumption of $\mu_{y} \geq 0$. Hence, the only solution where the $\mu_{y}$ satisfies the condition would be $\mu_{y} = 0 \; \forall \; y$. Fig. 1 shows the graphical representation of the constraint space and estimation of its gradients. In Fig. 1, it is observed that the directions of the gradient of $\mathbf{C} = \{\mathbf{C}1, \mathbf{C}2,..., \mathbf{C}n\}$ and $\mathbf{O}$ are opposite in direction. This is because if the complete KKT Lagrangian function is formulated,
\begin{equation}
\mathbf{O}x(z) = - \sum_{y=1}^{n} \mathbf{C}y(z)
\end{equation}

Eqn. $(5)$ symbolizes that the objective function (or gradient of the objective) is negative of the direction of constraint function (or its gradient). This means that $\triangledown_{z} \mathbf{O}$ is not found within the cone formed by the active constraint function space for which $\mu_{y} < 0 \Rightarrow \mu_{y} = 0$.

$\mathbf{Case \; 2}$: Consider the following OP,
\begin{equation}
\mathcal{P} \Rightarrow
\begin{cases}
\mathbf{min} \; \mathbf{O}x(z) \Rightarrow \mathbf{O}x(z) > 0 \; \forall \; z, x=1,...,m \\
\mathbf{C}y(z) \leq \mathbf{W}y, \mathbf{W} > 0 \\
\end{cases}
\end{equation}

Since $(5)$ is a minimization problem, formulating the KKT Lagrangian and estimating the regularity condition,
\begin{equation}
\mathcal{L}(z, \mu_{y}) = \mathbf{O}x(z) + \mu_{y} (\mathbf{C}y(z) -  \mathbf{W}y) \; \forall \; x, y
\end{equation} 

where $\mathbf{W}y$ is the upper bound on the constraints.
\begin{equation}
\underbrace{\triangledown_{z} \mathbf{O}x(z)}_\text{$> 0$} + \mu_{y} \triangledown_{z} (\underbrace{\mathbf{C}y(z)}_\text{$> 0$}) = 0
\end{equation} 
\begin{equation}
\mu_{y} = \frac{\triangledown_{z} \mathbf{O}x(z)}{\underbrace{- \triangledown_{z} \mathbf{C}y(z)}_\text{$\triangledown_{z} \mathbf{C}y(z)$}} = 0
\end{equation} 

Fig. 2(a) shows the graphical interpretation of the constraint space, objective function and its gradients for different upper bounds, i.e., $a, b,...,w$. In Fig. 2(a), $\triangledown_{z} \mathbf{O}$ is also not found within the cone formed by the constraint function space for which $\mu_{y} = 0$ holds true.

$\mathbf{Case \; 3}$: Consider the following OP,
\begin{equation}
\mathcal{P}1 \Rightarrow
\begin{cases}
\mathbf{max} \; \mathbf{O}x(z) \Rightarrow \mathbf{O}x(z) > 0 \; \forall \; z, x=1,...,m \\
\mathbf{C}y(z) \leq \mathbf{W}y  \Rightarrow \mathbf{C}y(z) > 0 \; \forall \; z, y=1,...,n \\
\end{cases}
\end{equation}

Formulating the KKT Lagrangian and estimating its regularity condition,
\begin{equation}
\mathcal{L}(z, \mu_{y}) = \mathbf{O}x(z) \underbrace{-}_\text{$\mathbf{max \; OP}$} \mu_{y} (\mathbf{C}y(z) - \mathbf{W}y) \; \forall \; x, y
\end{equation} 
\begin{equation}
\underbrace{\triangledown_{z} \mathbf{O}x(z)}_\text{$> 0$} - \mu_{y} \triangledown_{z} (\underbrace{\mathbf{C}y(z)}_\text{$> 0$}) = 0
\end{equation} 

\begin{equation}
\mu_{y} = \frac{\triangledown_{z} \mathbf{O}x(z)}{\triangledown_{z} \mathbf{C}y(z)} > 0
\end{equation} 

\begin{equation}
\mathcal{P}2 \Rightarrow
\begin{cases}
\mathbf{min} \; \mathbf{O}x(z) \Rightarrow \mathbf{O}x(z) > 0 \; \forall \; z, x=1,...,m \\
\mathbf{C}y(z) \geq 0  \Rightarrow \mathbf{C}y(z) > 0 \; \forall \; z, y=1,...,n \\
\end{cases}
\end{equation} 

Re-expressing the constraint as $(- \mathbf{C}y(z) \leq 0)$. Formulating the KKT Lagrangian and estimating the regularity condition,
\begin{equation}
\mathcal{L}(z, \mu_{y}) = \mathbf{O}x(z) \underbrace{+}_\text{$\mathbf{min \; OP}$} \mu_{y} (- \mathbf{C}y(z)) \; \forall \; x, y
\end{equation} 
\begin{equation}
\underbrace{\triangledown_{z} \mathbf{O}x(z)}_\text{$> 0$} - \mu_{y} \triangledown_{z} (\underbrace{\mathbf{C}y(z)}_\text{$> 0$}) = 0
\end{equation} 
\begin{equation}
\mu_{y} = \frac{\triangledown_{z} \mathbf{O}x(z)}{\triangledown_{z} \mathbf{C}y(z)} > 0
\end{equation} 

Fig. 2(b) and 3 shows the graphical interpretation of the constraint space, objective function and its gradients for $\mathcal{P}1$ and $\mathcal{P}2$. If one formulates the KKT Lagrangian of $\mathcal{P}1$ or $\mathcal{P}2$,
\begin{equation}
\mathbf{O}x(z) = \sum_{y=1}^{n} \mathbf{C}y(z) 
\end{equation}

Eqn. $(18)$ shows that the objective function sis estimated as the summation of $n$ non-negative constraint functions. This symbolizes that the gradient of $\mathbf{O}x(z)$ will be found within the cone formed by the active constraints as shown in Fig. 3 for which $\mu_{y} > 0$. In all the above cases presented, $\mu_{y}$ is estimated where $\mathbf{O}x(z), \mathbf{C}y(z) > 0$ $\forall \; x, y, z$. Table I highlights the KKT multiplier $\mu_{y}$ for other cases. \\ \\
$\mathbf{Extended \; result \; 1}$: Consider the following OP,
\begin{equation}
\mathcal{P} \Rightarrow
\begin{cases}
\mathbf{max} \; \mathbf{O}x(z) \Rightarrow \mathbf{O}x(z) > 0 \; \forall \; z, x=1,...,m \\
\mathbf{C}y(z) \geq 0 \; \mathbf{for} \; y = 1 \\
\mathbf{C}y(z) \leq \mathbf{W}y \; \mathbf{for} \; y = 2,...,n \\
\end{cases}
\end{equation}
Formulating the KKT Lagrangian with regularity condition,
\begin{equation}
y = 1 \Rightarrow \mathbf{O}x(z) + (\mu_{y} \mathbf{C}y(z)) = 0
\end{equation}
\begin{equation}
y = 2,...,n \Rightarrow \mathbf{O}x(z) - \sum_{y=2}^{n} (\mu_{y} \mathbf{C}y(z) - \mathbf{W}y) = 0
\end{equation}

Fig. 4 shows the graphical representation for estimation of $\mu_{y}$. It is observed that the direction of gradient of $\mathbf{O}$ and $\mathbf{C}y, y = 2,...,n$ is the same, i.e., the objective function is within the cones formed by the constraint space. This leads to $\mu_{y} > 0, y = 2,...,n$. On the other hand, the direction of gradient of $\mathbf{O}$ and $\mathbf{C}1$ are opposite for which $\mu_{y} = 0$. \\ 
Similarly, for a minimization OP: $\mathbf{min} \; \mathbf{O}x(z)$,\\
$\mathbf{Extended \; result \; 2}$: 
\begin{equation}
y = 1 \Rightarrow \mathbf{O}x(z) + (\mu_{y} (-\mathbf{C}y(z))) = 0
\end{equation}
\begin{equation}
y = 2,...,n \Rightarrow \mathbf{O}x(z) + \sum_{y=2}^{n} (\mu_{y} \mathbf{C}y(z) - \mathbf{W}y) = 0
\end{equation}

Fig. 5 shows the graphical estimation of KKT Lagrangian multipliers. As seen, for $y > 1$, $\mu_{y} = \frac{\triangledown_{z} \mathbf{O}x(z)}{\underbrace{- \triangledown_{z}  \mathbf{C}y(z)}_\text{$\triangledown_{z}  \mathbf{C}y(z)$}}$, i.e., the direction of gradient of $\mathbf{O}$ and $\mathbf{C}y$ are opposite. However, for $y = 1$, $\mu_{y} = \frac{\triangledown_{z} \mathbf{O}x(z)}{\triangledown_{z}  \mathbf{C}y(z)}$ for which $\mu_{y} > 0, y = 1$.

\section{Result 2}

\textbf{Result 2}: \textit{Estimation of the utility convergance multiplier} $\beta$ \textit{for the objective functions in the MOP}.

An important problem in MOP is defining a cost (or error) function by combining the optimization objectives using a scalar. The aim is to minimize the trade-offs across the objective functions, i.e., minimize the cost (or error) function over a defined variable. Let $\mathbf{Ox}(r_{i})$ $\forall \; x = [1, n]$ represent $n$ non-negative non-linear objectives over a resource (variable) $r_{i}$. A cost function is defined as a linear combination of the objectives,
\begin{equation}
\mathbf{E}(\beta, r_{i}) = \sum_{x = 1}^{n-1} \beta_{x} \mathbf{Ox}(r_{i}) + (\underbrace{1 -  \sum_{x = 1}^{n-1} \beta_{x}}_\text{$\beta_n$}) \mathbf{On}(r_{i}), 0 \leq \beta_{x} \leq 1
\end{equation}
\hspace{0.1cm}
where $\beta = \{\beta_1, \beta_2,...,\beta_n\}$ are the scalars across $n$ objective functions. For a given $\beta_{x}, 1 \leq x \leq n$,  $\exists \; r_{i}$ which minimizes the cost function $\mathbf{E}(\beta,r_{i})$. We define,
\begin{equation}
\mathbf{E}^{*}(\beta_x) = \mathbf{min}_{r_{i}^{*}} \; \mathbf{E}(\beta_x, r_{i}) = \mathbf{E}(\beta_x, r_{i}^{*} (\beta_x))
\end{equation}

where $r_{i}^{*}(\beta_x) = r_{i}^{*}$ is the resource variable estimated at optimum $\beta_x$. Eqn. $(25)$ should be convex, irrespective of the objective functions convexity. Let $\mathbf{E}_{x}^{*} = \mathbf{E}^{*}(\beta_{x})$. Then, for any $0 \leq \alpha \leq 1$, from the concept of convexity, \\ \\
$\mathbf{E}^{*}(\underbrace{\sum_{x=1}^{n-1} \alpha_{x} \beta_{x}  + (1 - \sum_{x=1}^{n-1} \alpha_{x}) (1 - \sum_{x = 1}^{n-1} \beta_{x})}_\text{$ \alpha_{1} \beta_{1} +  \alpha_{2} \beta_{2} + ... +  \alpha_{n} \beta_{n}$} ) $
\begin{equation}
\begin{split}
& =  \mathbf{min}_{r_{i}}\mathbf{E} (\sum_{x=1}^{n-1} \alpha_{x} \beta_{x}  + (1 - \sum_{x=1}^{n-1} \alpha_{x}) (1 - \sum_{x = 1}^{n-1} \beta_{x}), r_{i}) \\ &  \leq \sum_{x=1}^{n-1} \alpha_{x} \; \underbrace{\mathbf{min}_{r_i} \; \mathbf{E} (\sum_{x=1}^{n-1} \beta_{x}, r_i)}_\text{$\mathbf{E}^{*}(\beta_{x})$} + (1 - \sum_{x=1}^{n-1} \alpha_{x}) \; \\ & \underbrace{\mathbf{min}_{r_i} \; \mathbf{E} (\underbrace{1 - \sum_{x=1}^{n-1} \beta_{x}}_\text{$\beta_n$}, r_i)}_\text{$\mathbf{E}^{*}(\beta_{n})$}
\end{split}
\end{equation}
\hspace{0.1cm}
It may be assumed without loss of generality that $\mathbf{Ox}(r_{i}) = 0, x > 1$ for certain $r_{i}$, such that $\mathbf{O1}(r_{i}) \neq 0$. Then, $(25) \Rightarrow$
\begin{equation}
\begin{split}
\mathbf{E}^{*} (\beta) & = \mathbf{min}_{r_i} \; \{\beta [\mathbf{O1}(r_i)]\} 
\end{split}
\end{equation}
\hspace{0.1cm}
So far, it has been observed that $\mathbf{E}^{*} (\beta)$ had been minimized over $r_{i}$. If $\mathbf{E}^{*} (\beta)$ is further minimized over $\beta$, the cost function might become too low. One way to ensure that $\mathbf{E}^{*} (\beta)$ is not too low such that $\mathbf{E}^{*} (\beta)$ is within a certain range is to now find the maximum cost function with respect to $\beta$ (since $r_{i}$ has been already used in the minimization operation). This symbolizes that the operations have been performed considering both extremes (min, max) to ensure that $\mathbf{E}^{*} (\beta)$ is within the range.

Let $\beta^{*}$ be the weight that now maximizes $\mathbf{E}^{*} (\beta)$. Previously, $\mathbf{E}^{*} (\beta)$ has been minimized over $r_i$. Let $\beta^{*}$ be the weight that now maximizes $\mathbf{E}^{*} (\beta)$ $\implies$ $r_{i}^{*} = r_{i}(\beta^{*})$. Then,  
\begin{equation}
\begin{split}
\frac{\partial (\mathbf{E}^{*} (\beta))}{\partial \beta} & =  \frac{\partial (\mathbf{E} (\beta, r_{i}^{*}))}{\partial \beta} + \underbrace{\frac{\partial (\mathbf{E} (\beta^{*}, r_{i}))}{\partial r_{i}}}_\text{$= 0$}  \frac{d (r_{i}^{*})}{d \beta} = 0
\end{split}
\end{equation}
\hspace{0.1cm}
$\frac{\partial (\mathbf{E} (\beta^{*}, r_{i}))}{\partial r_{i}} = 0$ since $r_{i}$ minimizes the cost function $\implies$ 
\begin{equation}
\frac{\partial (\mathbf{E} (\beta, r_{i}^{*} (\beta)))}{\partial \beta}|_{\beta = \beta^{*}} = 0 \Rightarrow \frac{\partial (\beta [\mathbf{O1}(r_i)])}{{\partial \beta}} = 0
\end{equation}

$\Rightarrow \mathbf{O1}(r_i)$ = 0 or $\mathbf{O1}(r_{i}^{*}) = 0$ for $r_{i}^{*}$ estimated at $\beta$. Since $\nexists \; r_{i}^{*}$, such that $\mathbf{O1}(r_{i}^{*}) = 0$ and $\mathbf{O1} > 0$ always, $\mathbf{O1}$ is independent of $\beta$. Hence, from $(27)$ $\xRightarrow{\beta = 1}$
\begin{equation}
\mathbf{E}^{*}(1) =  \mathbf{min} \; \mathbf{O1}(r_{i}^{*}) \neq 0
\end{equation}

Though $(27)$ might not seem to be a typical \textbf{minimization (or trade-off)} operation on the cost function (due to a single objective function), it has been proved (mathematically) that $\mathbf{E}^{*}$ and $\mathbf{O1}$ are independent of $\beta$ as $\beta = 1$.

\end{document}